\documentclass{aa}  

\usepackage{graphicx}
\usepackage{txfonts}
\usepackage{natbib}
\usepackage{hyperref}
\usepackage{tabularx,booktabs}
\newcommand{\kms}{\, \rm{km \, s^{-1}}}
\newcommand{\micron}{\,\mu\rm{m}}
\newcommand{\arcs}{^{\prime\prime}}
\newcommand{\angstrom}{\mbox{\normalfont\AA}}

\begin{document}

   \title{Infrared view of the multiphase ISM in NGC\,253}

   \subtitle{I. Observations and fundamental parameters of the ionised gas}

   \author{A. Beck
          \inst{1,2}
          \and
          V. Lebouteiller\inst{2}
          \and
          S. C. Madden\inst{2}
          \and
          C. Iserlohe\inst{1}
          \and
          A. Krabbe\inst{1}
          \and
          L. Ramambason\inst{2}
          \and
          C. Fischer\inst{1}
          \and
          M. Kaźmierczak-Barthel\inst{1}
          \and
          S. T. Latzko\inst{1}
          \and
          J. P. Pérez-Beaupuits\inst{3,4}
          }

   \institute{Deutsches SOFIA Institut, University of Stuttgart,
              Pfaffenwaldring 29, D-70569 Stuttgart\\
              \email{abeck@dsi.uni-stuttgart.de}
         \and
             Laboratoire AIM, CEA/Service d'Astrophysique, Bât. 709, CEA-Saclay, 91191, Gif-sur-Yvette Cedex, France
        \and
            Max-Planck-Institut für Radioastronomie, Auf dem Hügel 69, 53121 Bonn, Germany
        \and
            Centro de Astro-Ingeniería (AIUC), Pontificia Universidad Católica de Chile, Av. Vicuña Mackena 4860, Macul, Santiago, Chile
		\\
			}
   \date{}

 
  \abstract
  {Massive star-formation leads to enrichment with heavy elements of the interstellar medium. On the other hand, the abundance of heavy elements is a key parameter to study the star-formation history of galaxies. Furthermore, the total molecular hydrogen mass, usually determined by converting CO or $\left[\ion{C}{ii}\right]158\micron$ luminosities, depends on the metallicity as well. The excitation of metallicity-sensitive emission lines, however, depends on the gas density of \ion{H}{ii} regions, where they arise.}  
   {We used spectroscopic observations from SOFIA, \textit{Herschel}, and \textit{Spitzer} of the nuclear region of the starburst galaxy NGC\,253, as well as photometric observations from GALEX, 2MASS, \textit{Spitzer}, and \textit{Herschel} in order to derive physical properties such as the optical depth to correct for extinction, as well as the gas density and metallicity of the central region.}
   {Ratios of the integrated line fluxes of several species were utilised to derive the gas density and metallicity. The $\left[\ion{O}{iii}\right]$ along with the $\left[\ion{S}{iii}\right]$ and $\left[\ion{N}{ii}\right]$ line flux ratios for example, are sensitive to the gas density but nearly independent of the local temperature. As these line ratios trace different gas densities and ionisation states, we examined if these lines may originate from different regions within the observing beam. The $(\left[\ion{Ne}{ii}\right]13\micron + \left[\ion{Ne}{iii}\right]16\micron)$/Hu $\alpha$ line flux ratio on the other hand, is independent of the depletion onto dust grains but sensitive to the Ne/H abundance ratio and will be used as a tracer for metallicity of the gas.}
   {We derived values for gas phase abundances of the most important species, as well as estimates for the optical depth and the gas density of the ionised gas in the nuclear region of NGC\,253. We obtained densities of at least two different ionised components ($<84\,\mathrm{cm}^{-3}$ and $\sim 170-212\,\mathrm{cm}^{-3}$) and a metallicity of solar value.}
   {}


   \maketitle
%
\section{Introduction}
NGC\,253 is the nearest spiral galaxy showing a nuclear starburst with a star-formation rate (SFR) of $3\,\mathrm{M}_{\sun}\,\mathrm{yr}^{-1}$ \citep{Radovich2001}.
Its proximity makes this galaxy a perfect model to study the interstellar medium (ISM) in the extreme environments of massive star-formation.
Near-infrared observations revealed the presence of a bar which funnels gas and dust from outer regions into the nuclear starburst \citep{Iodice2014}.
Besides fuelling of the starburst by the stellar bar, an outflow exiting from the nuclear region has been observed in H$\alpha$ \citep{Bolatto2013}, and CO \citep{Leroy2015, Walter2017}.
From these observations, a mass outflow rate of about $3-9\,\mathrm{M}_{\sun}\,\mathrm{yr}^{-1}$ has been estimated, implying that the nuclear starburst is severely suppressed by the outflow.
The existence of a central supermassive black hole has been discussed within literature \citep[e.g. ][]{Vogler1999,Guenthardt2015}.
The putative central supermassive black hole and subsequent active galactic nucleus \citep[AGN, ][]{FernandezOntiveros2009} should have a noticeable impact on the ISM and the star-formation activity in the nuclear region, by accreting the surrounding stars and ISM at high velocities.
The accretion creates highly ionised species such as O$^{3+}$ or Ne$^{4+}$, which are also observable via their fine-structure lines in the mid-infrared \citep[MIR, ][and references therein]{FernandezOntiveros2016}.
The outflows originating from the starburst and the hypothetical AGN reduces the amount of the cold molecular gas reservoir, which is needed to sustain the ongoing starburst.
The nature of this hypothetical black hole, however, its signatures in the surrounding ISM and consequences for star-formation and the H$_{2}$ gas mass reservoir are still undetermined \citep{FernandezOntiveros2009}.
Due to the presence of this many different components $-$ bar \citep{Engelbracht1998}, starburst \citep{Rosenberg2013}, outflows \citep{Bolatto2013,Krips2016}, and possibly a black hole or AGN \citep{FernandezOntiveros2009} $-$ the major energy input to the ISM is uncertain and will be tackled in this series.

\begin{table}
    \centering
    \caption{General properties of NGC\,253.}
    \begin{tabular}{lcl}
        \toprule
        \toprule
        Quantity & Value & Reference \\
        \midrule
        RA (J2000) & $0\mathrm{h}47\mathrm{m}33.1\mathrm{s}$ & \citet{Mueller2010} \\
        Dec (J2000) & $-25^{\circ}17\arcmin18\arcsec.3$ & \citet{Mueller2010} \\
        Distance & $3.5\,\mathrm{Mpc}$ & \citet{Rekola2005} \\
        Image Scale & $17\,\mathrm{pc}/\arcsec$ &  \\
        SFR & $3\,\mathrm{M}_{\sun}\,\mathrm{yr}^{-1}$ & \citet{Radovich2001} \\
        Inclination & $78^{\circ}$ & \citet{Pence1981} \\
        $v_{\mathrm{sys}}$ & $236\kms$ & \citet{Pence1981} \\
        \bottomrule
    \end{tabular}
    \tablefoot{Right ascension (RA) and Declination (Dec) of the nucleus. Distance calculated from planetary nebula luminosity function. The SFR was calculated from ISO FIR observations, $v_{\mathrm{sys}}$ is the heliocentric systemic velocity of the source.}
    \label{tab:N253properties}
\end{table}

Additionally, there is still little knowledge about the chemical composition of the gas in the nuclear region of NGC\,253 \citep[][]{Martin2021}.
This is mostly due to difficulties that occur when optical lines like $\left[ \ion{O}{ii} \right]3737\angstrom$ or $\left[ \ion{O}{iii} \right]5007\angstrom$ are observed in such a highly extincted starburst galaxy.
The metallicity, however, is a key parameter when trying to understand the ongoing mechanisms in the ISM.
Since metals are by-products of star-formation, the metallicity of a galaxy tracts its evolutionary history.
Additionally, to estimate the molecular hydrogen gas mass reservoir of galaxies from CO emission, the conversion factor is known to be a function of metallicity \citep[e.g. ][and references therein]{Bolatto2013ARAA}.
Recent studies showed that, in low-metallicity environments in particular, a large fraction of H$_{2}$ is not traced by CO emission \citep[e.g. ][and references within]{Madden2020}.
Hence, knowing the metallicity is the main parameter to estimate the molecular hydrogen mass, which itself is the key ingredient for star-formation.

This study uses recent SOFIA observations of $\left[ \ion{O}{iii} \right]52,\,88\micron$, $\left[ \ion{O}{i} \right]63,\,146\micron$, and $\left[ \ion{C}{ii} \right]158\micron$ in conjunction with \textit{Spitzer} and \textit{Herschel} mid-infrared (MIR) and far-infrared (FIR) spectroscopy.
We gather MIR to FIR observations from the various telescopes, some observations of which will be used in this study.
In this paper we address the properties of the ionised gas in the central region of NGC\,253, using Hu $\alpha\,(7-6)$, $\left[ \ion{Ne}{ii} \right]13\micron$, $\left[ \ion{Ne}{iii} \right]16\micron$, $\left[ \ion{S}{iii} \right]19\micron$, $\left[ \ion{S}{iii} \right]33\micron$, $\left[ \ion{O}{iii} \right]52\micron$, $\left[ \ion{O}{iii} \right]88\micron$, $\left[ \ion{N}{ii} \right]122\micron$, and $\left[ \ion{N}{ii} \right]205\micron$.
Due to their wide range of critical densities and ionisation potentials, we are able to pin down the densities traced by these different emission lines.
Additionally, we quantify the abundances of various species, and thus the overall metallicity of the central region, since ratios like O/H and N/O provide measures for the star-formation history of a galaxy \citep[e.g. ][]{PerezMontero2013,Spinoglio2022}.
In a followup paper we present a more complex multiphase model using MULTIGRIS \citep{Lebouteiller2022,Ramambason2022} a Bayesian method determining probability density distributions of parameters of a multisector Cloudy \citep{Ferland2017} grid from a set of emission lines.
With this approach we will constrain parameters of the ISM such as intensity of the radiation field of different components of the ISM, temperature and luminosity of the putative black hole or AGN, and age of the starburst.

This work is structured as follows: 
Section 2 describes our observations and the data reduction, as well as archival data from different observatories, and the determination of line fluxes and errors.
Section 3 follows with an analysis using some of the observed lines, determining the density and metallicity of the ionised gas, as well as the optical depth of the nuclear region.
A summary and outlook of future work concludes this study in Sect. 4.

\begin{figure}
\centering
\resizebox{\hsize}{!}{\includegraphics[scale=1]{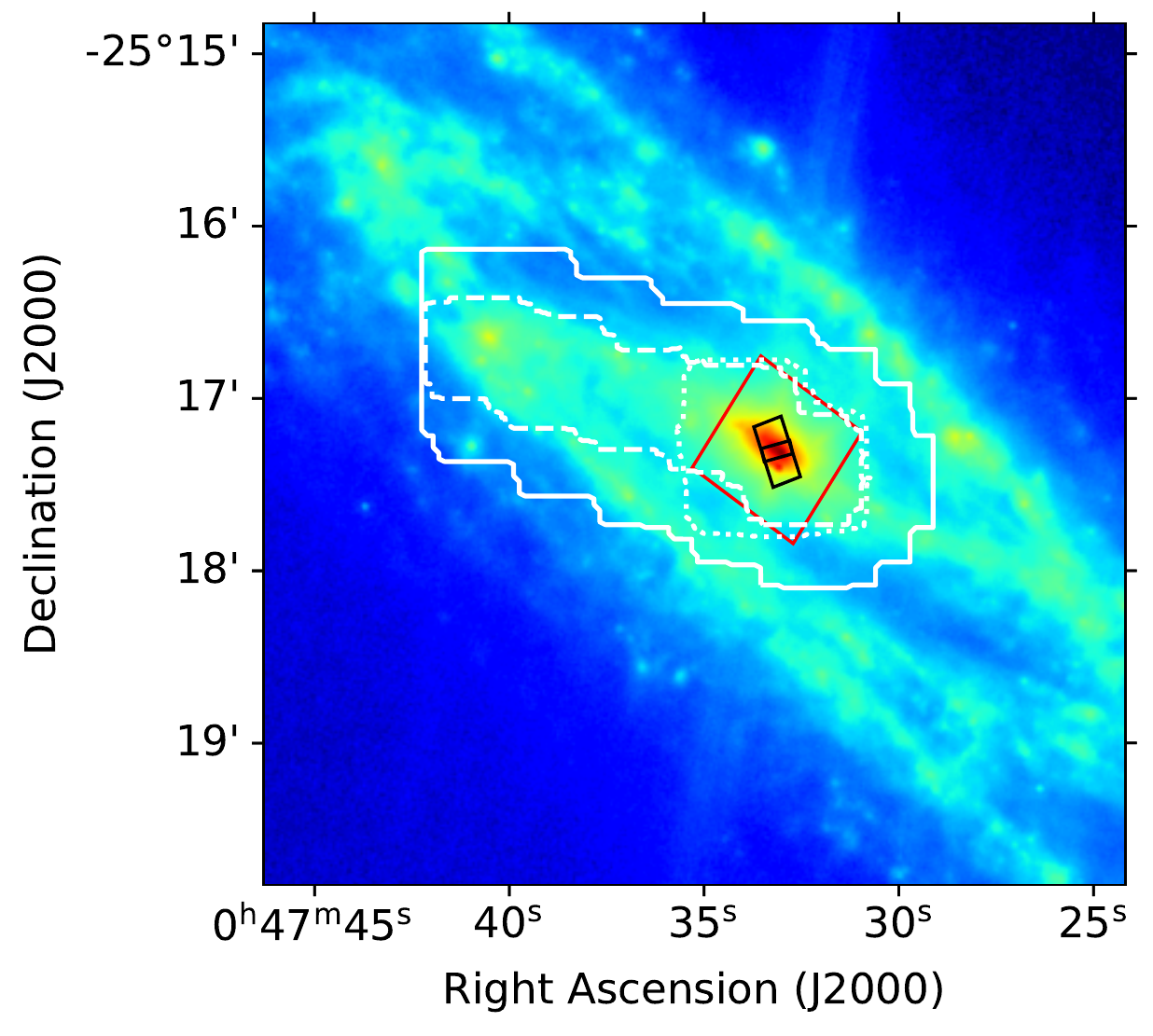}}
\caption{Photometric image from \textit{Spitzer}/IRAC at $\lambda=8\micron$ in logscale. The white contours show observations from SOFIA/FIFI-LS (solid - $\left[\ion{O}{i}\right]146\micron$ and $\left[\ion{C}{ii}\right]158\micron$, dashed - $\left[\ion{O}{iii}\right]88\micron$, dotted - $\left[\ion{O}{iii}\right]52\micron$ and $\left[\ion{O}{i}\right]63\micron$), the two black boxes show apertures of \textit{Spitzer}/IRS short- and long high modules. The red box shows the approximate footprint of the \textit{Herschel}/PACS observations.}
\label{fig:apertures}
\end{figure}


\section{Observations and data reduction}\label{sec:observations}
\subsection{SOFIA/FIFI-LS}\label{ssec:FIFILS}
We used the Field-Imaging Far-Infrared Line Spectrometer \citep[FIFI-LS][]{Fischer2018} onboard the Stratospheric Observatory For Infrared Astronomy (SOFIA, \citealt{Young2012}) to obtain emission line data.
FIFI-LS is an imaging spectrometer for the FIR providing a wavelength dependent spectral resolution of $R=600 - 2000$.
This instrument provides two independent observing channels allowing simultaneous observations in the $\lambda = 51 - 120\micron$ (blue channel) and in the $\lambda = 115 - 203 \micron$ (red channel) wavelength regimes.
Each channel projects an array of $5 \times 5$ spatial pixels (spaxels) on the sky, covering a total area of $30\arcs \times 30\arcs$ and $60\arcs \times 60\arcs$ in the blue and red channel, respectively \citep{Klein2014,Colditz2018}.

Observations were carried out on SOFIA flights starting from Palmdale, California during cycle 3 in October 2015, cycle 6 in November 2018, and cycle 7 in November 2019 as guaranteed time campaigns.
A summary of the observed lines, flight dates, and spatial and spectral resolution is listed in Table \ref{tab:observations}.
Figure \ref{fig:apertures} shows an outline of our observations (white contours).

\begin{table}
\centering
\caption{Summary of our observed lines with FIFI-LS.}
\begin{tabular}{lcrr}
\toprule
\toprule
Species & Flight Dates & $R$  & \multicolumn{1}{c}{PSF}  \\
   	    & 			   &      & \multicolumn{1}{c}{[$\arcs$]} \\
\midrule
$\left[\ion{O}{iii}\right]52\micron$  & 2015 Oct 14, 2018 Nov 07 & 980 & 5.5 \\
							  & 2018 Nov 08 &  &  \\
$\left[\ion{O}{i}\right]63\micron$ & 2015 Oct 14, 2015 Oct 23 & 1300 & 6.0 \\
$\left[\ion{O}{iii}\right]88\micron$ & 2015 Oct 23, 2018 Nov 07	 & 600 & 9.0 \\
 							 & 2019 Nov 02, 2019 Nov 05	 &  & \\
$\left[\ion{O}{i}\right]146\micron$ & 2015 Oct 23, 2018 Nov 07 & 1000 & 15.0 \\
							& 2018 Nov 08, 2019 Nov 05 &  &  \\
$\left[\ion{C}{ii}\right]158\micron$ & 2015 Oct 14, 2015 Oct 23 & 1150 & 16.0 \\
							 & 2019 Nov 02 & &  \\
\bottomrule
\end{tabular}
\tablefoot{Neutral and ionised species, observing flight dates, spectral and spatial resolution (FWHM of PSF) at the given wavelength of the observed lines. More details about spectral resolution can be seen in \citet{Fischer2016}}
\label{tab:observations}
\end{table} 

The FIFI-LS data were reduced with the FIFI-LS standard data reduction pipeline \citep{Vacca2020}.
The pipeline performs a flux calibration, and resamples the data onto a spatial grid of $1.5\arcs \times 1.5\arcs$ and $3\arcs \times 3\arcs$ (pipeline spaxels) in the blue and red channels, respectively.
The transmission of the atmosphere and hence the observed line flux at a given wavelength depend mainly on the total upward atmospheric precipitable water vapour (PWV).
Inflight PWV values are determined by targeting specific water vapour features which are used in the data reduction pipeline for atmospheric calibration purposes (see \citet{Fischer2021} and \citet{Iserlohe2021} for more details).
For our $\left[\ion{O}{iii}\right]52\micron$, $\left[\ion{O}{iii}\right]88\micron$, $\left[\ion{O}{i}\right]146\micron$, and $\left[\ion{C}{ii}\right]158\micron$ observations we used the data from the pipeline that is PWV calibrated. 
The $\left[\ion{O}{i}\right]63\micron$ lies in a wavelength regime that is contaminated with a telluric absorption feature and is therefore not used in this study.\\

\begin{figure}
        \centering
        \resizebox{\hsize}{!}{\includegraphics{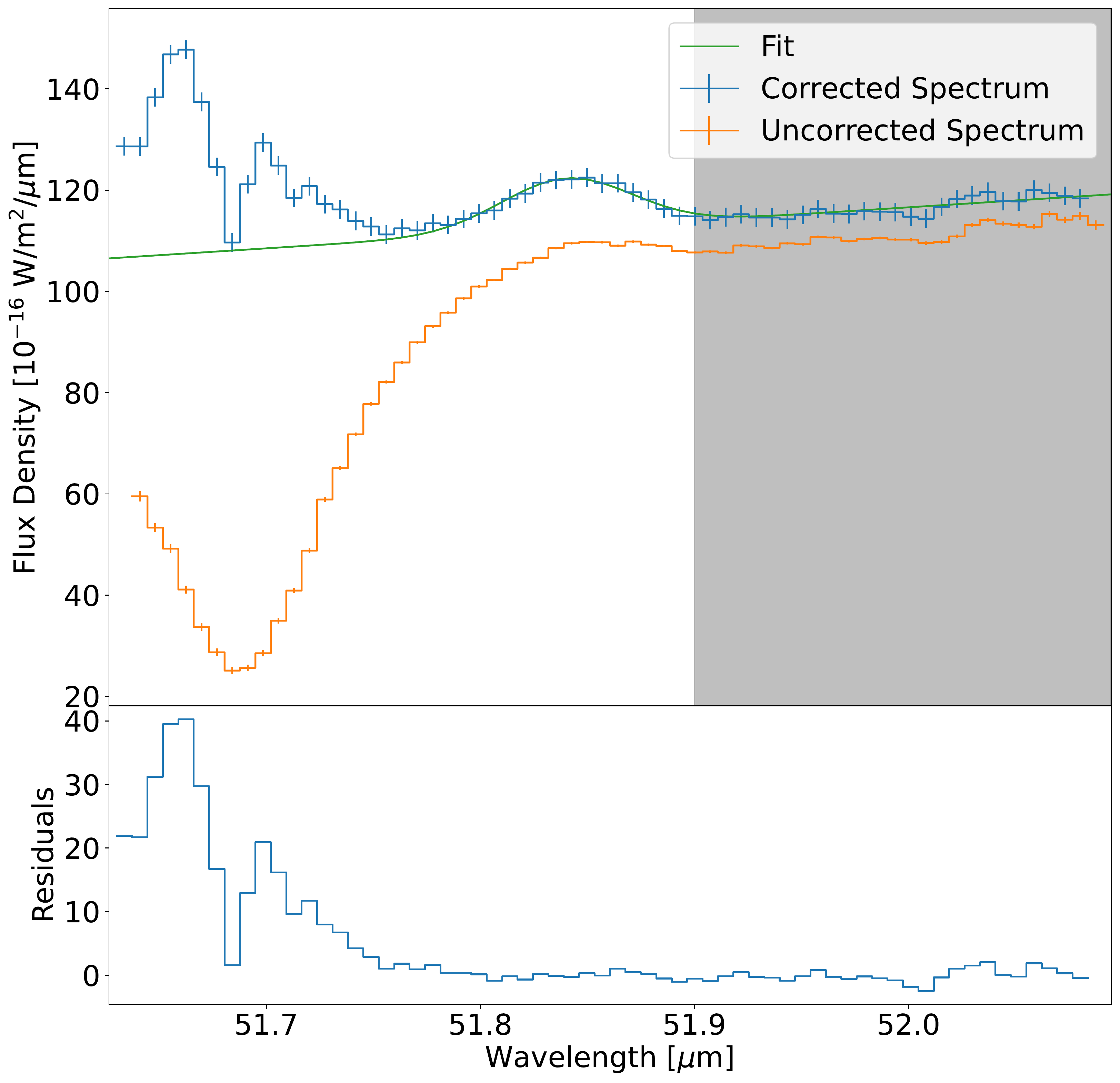}}
        \caption{Spectra of the pipeline spaxel ($1.5\arcs \times 1.5\arcs$) located toward the nucleus. The orange spectrum shows the observed spectrum (including errors from the ramp fit) without correction for atmospheric transmission. The telluric absorption feature at $51.7\micron$ can clearly be seen in the continuum. The blue spectrum shows the pipeline-corrected spectrum, including the errors determined as described in Sect. \ref{ssec:FIFILS}. The grey area shows the spectral region used to calculate flux density uncertainties. Green shows the best fit from the Monte-Carlo method. }
        \label{fig:52exampleSpec}
    \end{figure}

Flux density errors in the data cubes are the propagated errors from the ramp fit of the readout of the detector, but do not include errors from flux calibration or the PWV correction (FIFI-LS team, private communication).
We therefore estimated the statistical flux density uncertainty from the root-mean-square (RMS) error in the continuum region of each pipeline spaxel.
Due to the small spectral coverage the continuum is nearly constant and the RMS error is not amplified by a non-constant continuum.
For the $\left[\ion{O}{iii}\right]52\micron$, $\left[\ion{O}{iii}\right]88\micron$, $\left[\ion{O}{i}\right]146\micron$, and $\left[\ion{C}{ii}\right]158\micron$ observations we excluded spectral regions with the expected line position.
Due to a strong absorption feature at $51.7\micron$, the flux density $<51.8\micron$ was also not included for the root-mean-square calculation (see Fig. \ref{fig:52exampleSpec} for an example spectrum and fit).

\begin{figure*}
\centering
\includegraphics[width=17cm]{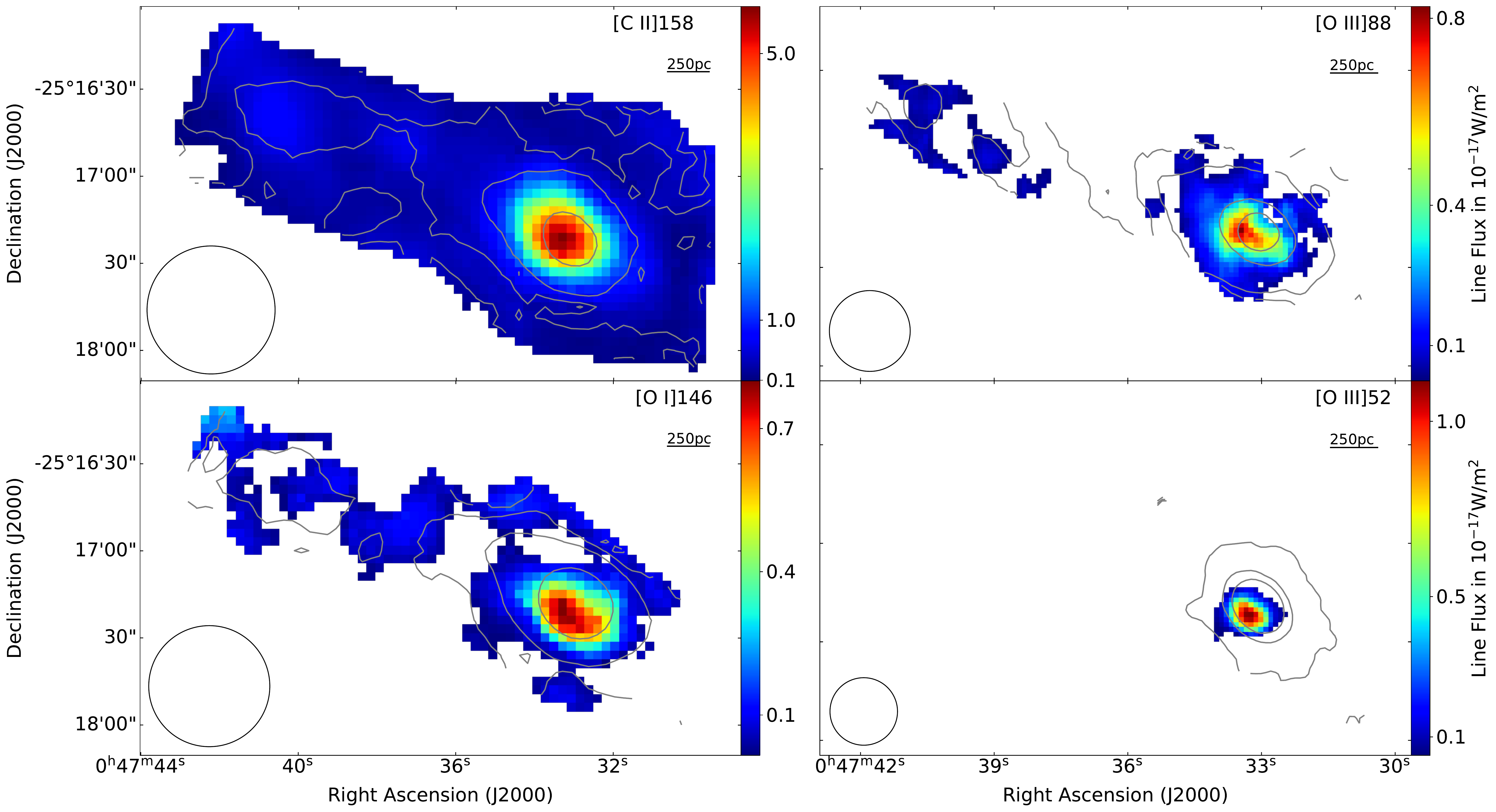}
\caption{Line flux maps from the SOFIA/FIFI-LS observations. Colourmaps show the line flux in $10^{-17}\, \rm{W \, m^{-2}}$ on a $3\arcs \times 3\arcs$ (left column) and $1.5\arcs \times 1.5\arcs$ (right column) grid, respectively. Pixels with a signal-to-noise ratio smaller than $3$ are masked. The black circles depict the aperture used to extract the spectrum from the nuclear region (Sect. \ref{sssec:fifi_hom}). We also show contours from the fitted underlying continuum (grey).}
\label{fig:fifiFlux}
\end{figure*}

To determine the line fluxes and their statistical uncertainties in each pipeline spaxel, we used a Monte-Carlo approach by varying the flux density within the normal-distributed flux density uncertainties in each spectral bin randomly over $500$ steps.
The spectrum obtained from this variation was fitted with a linear continuum and a Gaussian with variable mean, width and amplitude (see Fig. \ref{fig:52exampleSpec}).
From the integral over the Gaussian component we obtained the line flux in each step $-$ giving a sample of $500$ line fluxes in each pipelines spaxel, from which we calculated the mean and standard deviation, which are the line flux and line flux error, respectively.
These procedures were applied to the spectra from all pipeline spaxels, yielding the line flux maps shown in Fig. \ref{fig:fifiFlux}.

\subsection{Additional data}\label{ssec:ancillaryData}
\subsubsection{\textit{Spitzer}/IRS}
We complement our data set by single slit spectra obtained with the Infrared Spectrograph \citep[IRS,][]{Houck2004} onboard the Spitzer Space Telescope.
Data are available at the Combined Atlas of Spitzer IRS Spectra \citep[CASSIS\footnote{\url{https://cassis.sirtf.com/}}]{Lebouteiller2015} under AOR key 9072640. 

While the nuclear region of NGC\,253 was observed with low- and high-resolution modules, we only use the short-high (SH) and long-high (LH) module observations, from $9.6 - 19.6\micron$ and $18.7 - 27.2\micron$, since the low-resolution observations are saturated.
Both high-resolution modules have a velocity resolution of $\Delta v = 500\kms$, comparable to the spectral resolution of our FIFI-LS observations.
Slit sizes of the short- and long-wavelength module are different, $4.7\arcs \times 11.3\arcs$ and $11.1\arcs \times 22.3\arcs$, to better match the PSF size of each module.

CASSIS offers two reduction processes, the optimal extraction adapted to point-like sources, and full extraction adapted to extended sources.
Since the nucleus of NGC\,253 is partly extended (see Sect. \ref{sssec:irs_homogen} and Fig. \ref{fig:LH_spatialprofile}), we chose full extraction procedure as this ensures we retrieve most of the source's flux.
The full extraction accounts for losses of flux density, and integrates the flux density over the full slit size.
Although the nucleus of NGC\,253 is not uniformly extended over the whole slit, which would be the ideal application for the full extraction, this process yields better results than optimal extraction.

For each emission line in the IRS spectrum we fitted a Gaussian line profile and a polynomial of order 0 to 4 for the underlying continuum using \verb|lmfit|.
From the line fits, we obtained the line flux and line flux errors.
In cases where several emission lines were close together, we fitted the continuum and all lines simultaneously. 
The high order for the continuum is necessary to account for the several broad emission and absorption features from, for instance, polycyclic aromatic hydrocarbons and silicates in the MIR.
Line centroids are allowed to vary within $400\,\mathrm{km}\,\mathrm{s}^{-1}$ with respect to the expected velocity from the source (see Table \ref{tab:N253properties}).
The minimum line width is $300\,\mathrm{km}\,\mathrm{s}^{-1}$, which is about the spectral resolution of IRS.
Some lines are only weakly detected.
We give an upper limit for the line flux, if the obtained line flux from the fit is smaller than $2\times$ the RMS of the flux density of the continuum (in W/m$^{2}$/$\mu$m) times the spectral resolution (in $\mu$m).
In this case, the upper limit is given by $2\times$ the RMS error of the continuum.

\subsubsection{\textit{Herschel}}\label{sssec:PACS_fit}
We add to the FIFI-LS and \textit{Spitzer} data observations with the Photodetector Array Camera and Spectrometer \citep[PACS,][]{Poglitsch2010} onboard the Herschel Space Observatory\footnote{obs. ID 1342210652, 1342212531} \citep{Pilbratt2010}.
We used \textit{PACSman} v3.63 \citep{Lebouteiller2012} for data reduction and analysis of the PACS spectral cubes.
PACSman uses the rebinned data cubes from the PACS standard pipeline and fits a Gaussian emission line and a second order polynomial for the underlying continuum in each of the $9.4\arcs \times 9.4\arcs$ spaxels.
PACSman provides a Monte-Carlo based method to determine the line flux errors in each spaxel.
Similar to the procedure in Sect. \ref{ssec:FIFILS}, the flux density is varied within the Gaussian-distributed uncertainties in each of the $3000$ steps.
Each of the obtained spectra are fitted yielding a line flux in each spaxel.
From the sample of $3000$ line fluxes we calculated the mean and standard deviation, ultimately provides the line flux and line flux error in one spaxel, respectively.

\begin{figure*}
    \centering
    \includegraphics[width=17cm]{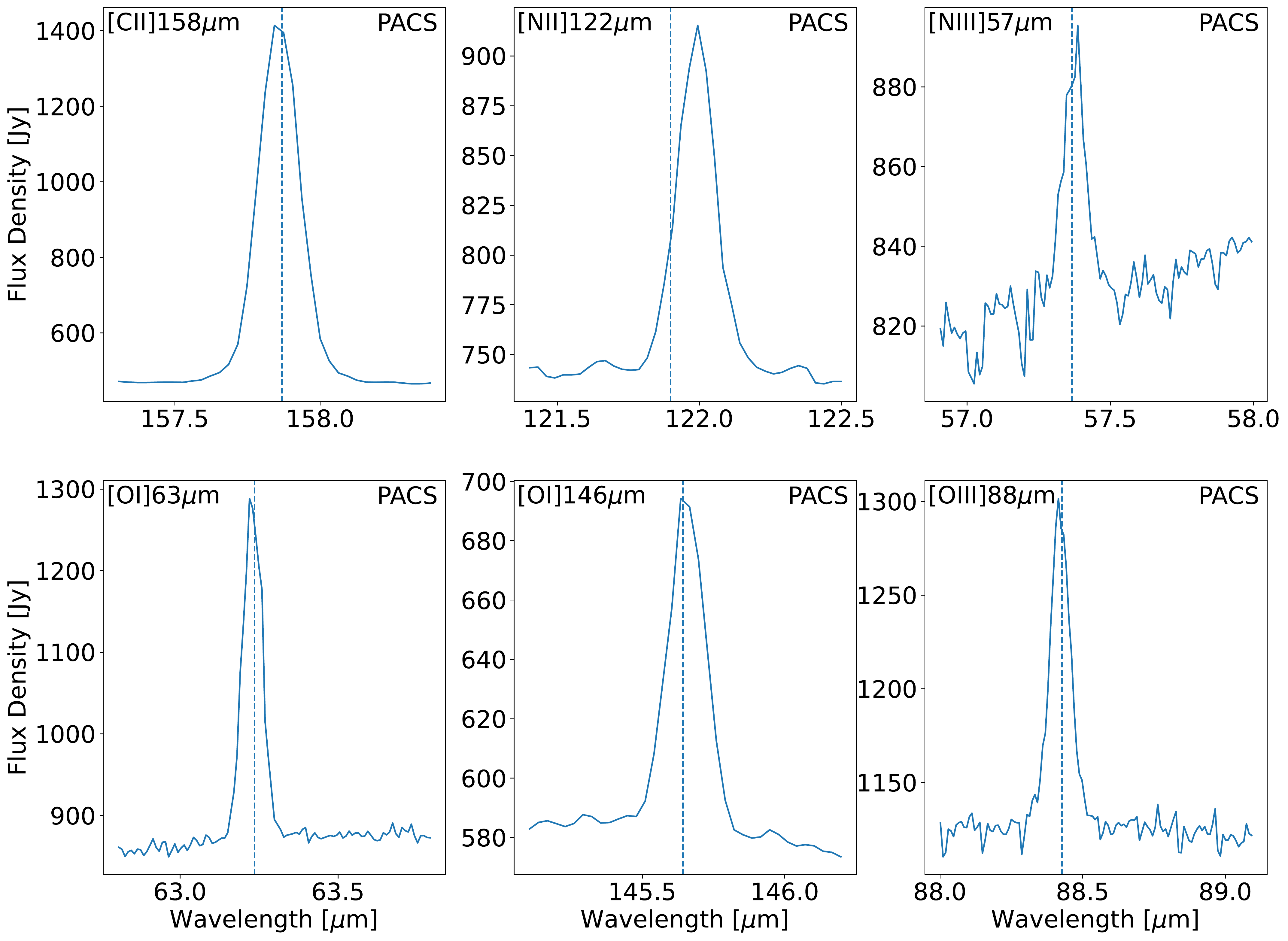}
    \caption{Spectra of the nuclear region observed with the \textit{Herschel}/PACS spectrometer. Only the spectrum of the central $3\times 3$ spaxels are shown here (Sect. \ref{sssec:pacs_hom}). The dashed line shows the redshift-corrected rest wavelength of the emission line.}
    \label{fig:PACS_spectra}
\end{figure*}

We obtained line flux maps and corresponding uncertainties for the $\left[\ion{N}{iii}\right]57\micron$, $\left[\ion{O}{i}\right]63\micron$, $\left[\ion{O}{iii}\right]88\micron$, $\left[\ion{N}{ii}\right]122\micron$, $\left[\ion{O}{i}\right]146\micron$, and $\left[\ion{C}{ii}\right]158\micron$ transitions.
See Fig. \ref{fig:PACS_spectra} for the observed spectra from the central $3\times 3$ spaxels of the field-integral unit (cf. Sect. \ref{sssec:pacs_hom}).

We finally complement our data set with \textit{Herschel}/SPIRE-FTS observations.
The SPIRE line fluxes from \citet{PerezBeaupuits2018} are already corrected for the semi-extended nuclear source (see Sect. \ref{ssec:homogenisation}), hence, we do not reduce these observations and list only the results in Table \ref{tab:spitzerIRS}.
For details of the data reduction see \citet{PerezBeaupuits2018}.

\subsection{Data homogenisation}\label{ssec:homogenisation}
We focus on emission from the nuclear source in this study.
Hence, we have to determine the line fluxes and line flux errors arising only from the nucleus for each instrument.

\subsubsection{SOFIA/FIFI-LS}\label{sssec:fifi_hom}
\begin{figure*}
    \centering
    \includegraphics[width=17cm]{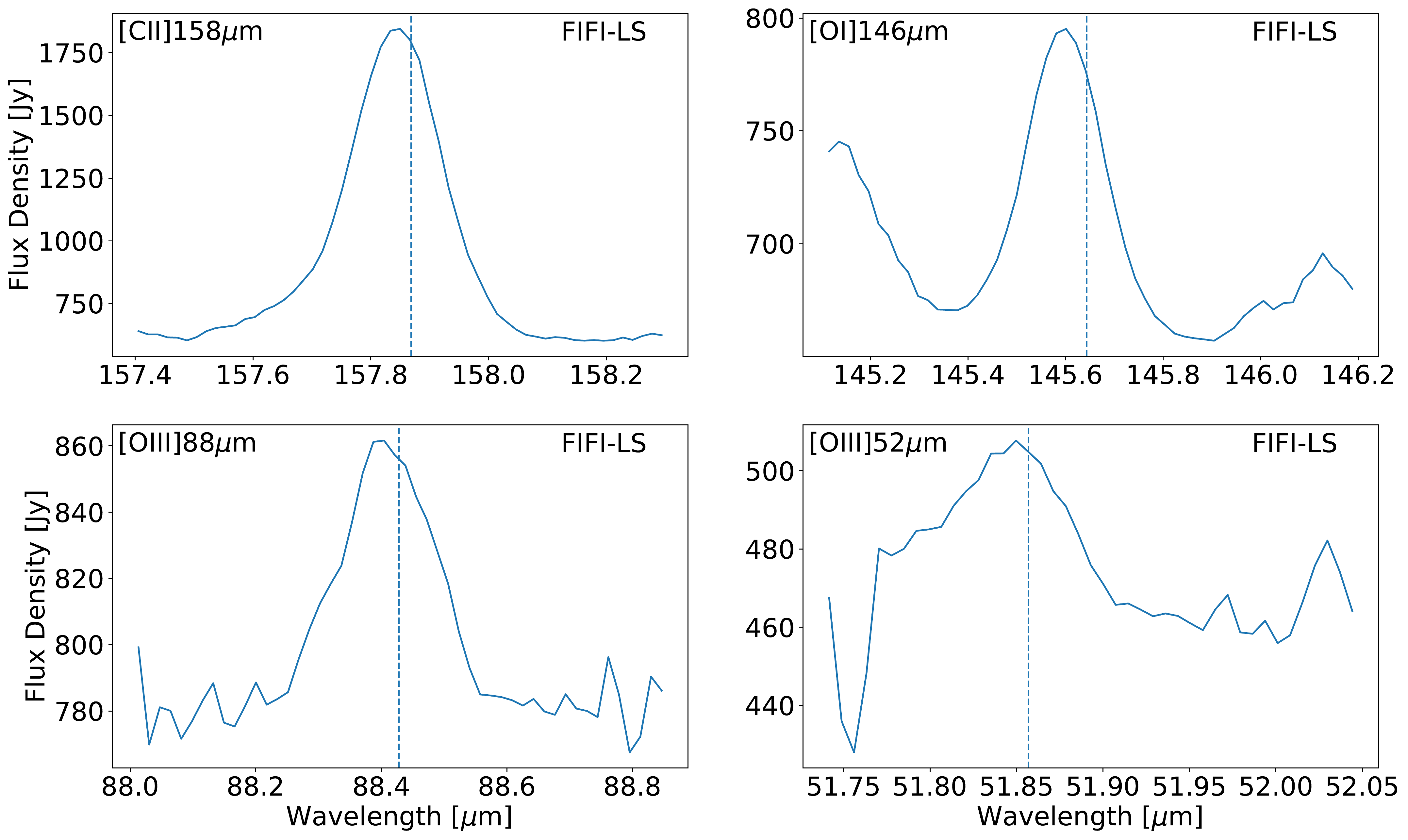}
    \caption{Spectra from our FIFI-LS observations in the nuclear region of NGC\,253. See Sect. \ref{sssec:fifi_hom} and Fig. \ref{fig:fifiFlux} for details of the extraction aperture.}
    \label{fig:FIFI_Spectra}
\end{figure*}

The FIFI-LS line flux maps (Fig. \ref{fig:fifiFlux}) are oversampled, without knowledge of the correlation between the pipeline spaxels.
In addition, the line flux maps are much more extended than only the nucleus and show emission from the bar and disk.
To extract line fluxes and corresponding errors from only the nucleus, we proceeded as follows:
we calculated an effective PSF size by fitting a Gaussian to the spatial profile of the nucleus for each emission line $-$ the ideal PSF is broadened by the non-Nyquist sampling of the observations and the resample procedure in the pipeline.
We resampled both, the line flux maps (Fig. \ref{fig:fifiFlux}) and line flux error maps to a spatial grid of the effective PSF size.
For the line flux error maps we assumed that all pixels are fully correlated $-$ meaning that the errors are cumulative.
To get most of the line flux from the nucleus, not contaminated from background (i.e. disk-) emission, we extracted the line flux and errors from the resampled maps using a circular aperture with a radius of $2\times$ the effective PSF FWHM, assuming that the pixels are uncorrelated.
This means that errors add up quadratically.
This procedure yields an upper limit for the line flux errors, since the assumption of the first step that all pixels are correlated is not entirely fulfilled.
However, it is a good approximation as the errors are of the same order as uncertainties from \textit{Herschel}/PACS (see Table \ref{tab:spitzerIRS}).
The resulting spectra from this procedure are shown in Fig. \ref{fig:FIFI_Spectra}. 

All line fluxes and corresponding statistical errors from the FIFI-LS observations are listed in the second block of Table \ref{tab:spitzerIRS}.
The line fluxes listed are  in good agreement with the values from \textit{Herschel}/PACS as explained in Sect. \ref{sssec:pacs_hom}).
We compared the $\left[\ion{O}{iii}\right]52\micron$ line flux with observations from the Kuiper Airborne Observatory (KAO), as this emission line was not observable with PACS.
Compared to observations published by \citet{Carral1994}, our value for $\left[\ion{O}{iii}\right]52\micron$ emission is lower by $\sim50\%$.
However, the beamsize of KAO at $52\,\mu$m is $\sim 8 \times$ larger than SOFIA at that wavelength, so the KAO observations most likely include some background emission.
To confirm that the KAO measurements of $\left[\ion{O}{iii}\right]52\micron$ include diffuse emission, we extracted the spectrum from a $43\arcs$ aperture, which is the beamsize of the KAO observations shown in \citet{Carral1994}.
Applying the same Monte-Carlo method as in Sect. \ref{ssec:FIFILS} we get a line flux of $107.7 \pm 30.8\,\mathrm{W} \, \mathrm{m}^{-2}$.
This is in good agreement with \citet{Carral1994}, who yielded a line flux of $90 \pm 20\,\mathrm{W} \, \mathrm{m}^{-2}$.

\subsubsection{\textit{Spitzer}/IRS}\label{sssec:irs_homogen}
\begin{figure}
        \centering
        \resizebox{\hsize}{!}{\includegraphics{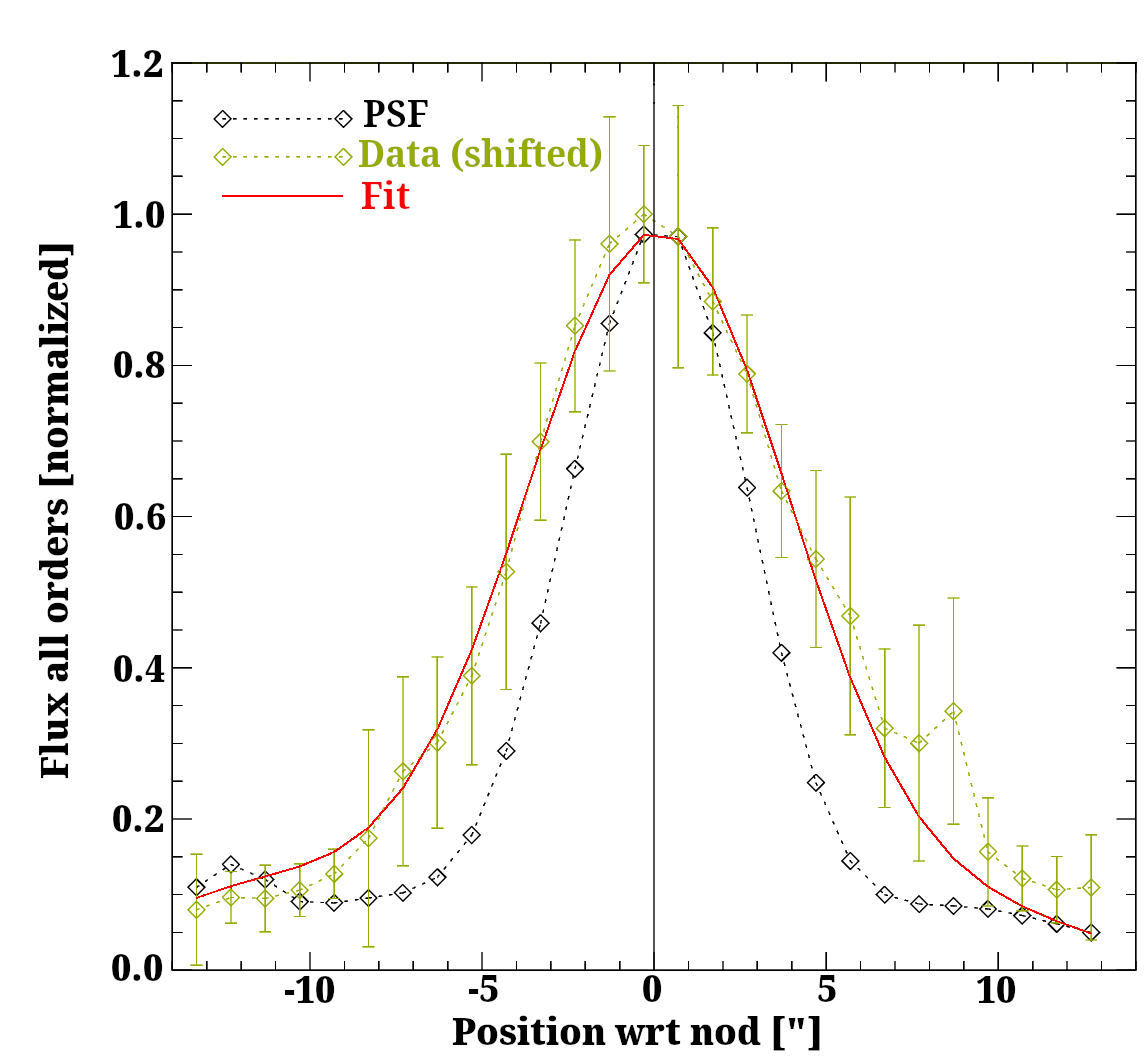}}
        \caption{Spatial profiles from the cross-dispersion direction of the LH module of IRS \citep[see][for details]{Lebouteiller2015}. Black shows the model PSF from observations of several point sources. Green diamonds show the spatial profile of NGC\,253 in the observations, red is a Gaussian fit with a free FWHM. From the fit we obtained a broadening of the PSF of $6.68\arcs$.}
        \label{fig:LH_spatialprofile}
    \end{figure}

The spatial profile of the nuclear source is not point-like, but slightly extended, as can be seen from the cross-dispersion direction of the IRS LH slit (see Fig. \ref{fig:LH_spatialprofile}).
From the comparison of the observed spread and the ideal PSF, we get a spatial extent of the nucleus  of $6.68\arcs$, corresponding to $113.6\,\mathrm{pc}$ under the assumed distance of $3.5\,\mathrm{Mpc}$ (see Table \ref{tab:N253properties}).
This is in good agreement with the $6\arcs$ from $\left[\ion{Ne}{ii} \right]13\micron$ observations by \citet{Engelbracht1998}.
We assume that the source size is the same for all MIR and FIR observations.

\begin{figure*}
\centering
\includegraphics[width=17cm]{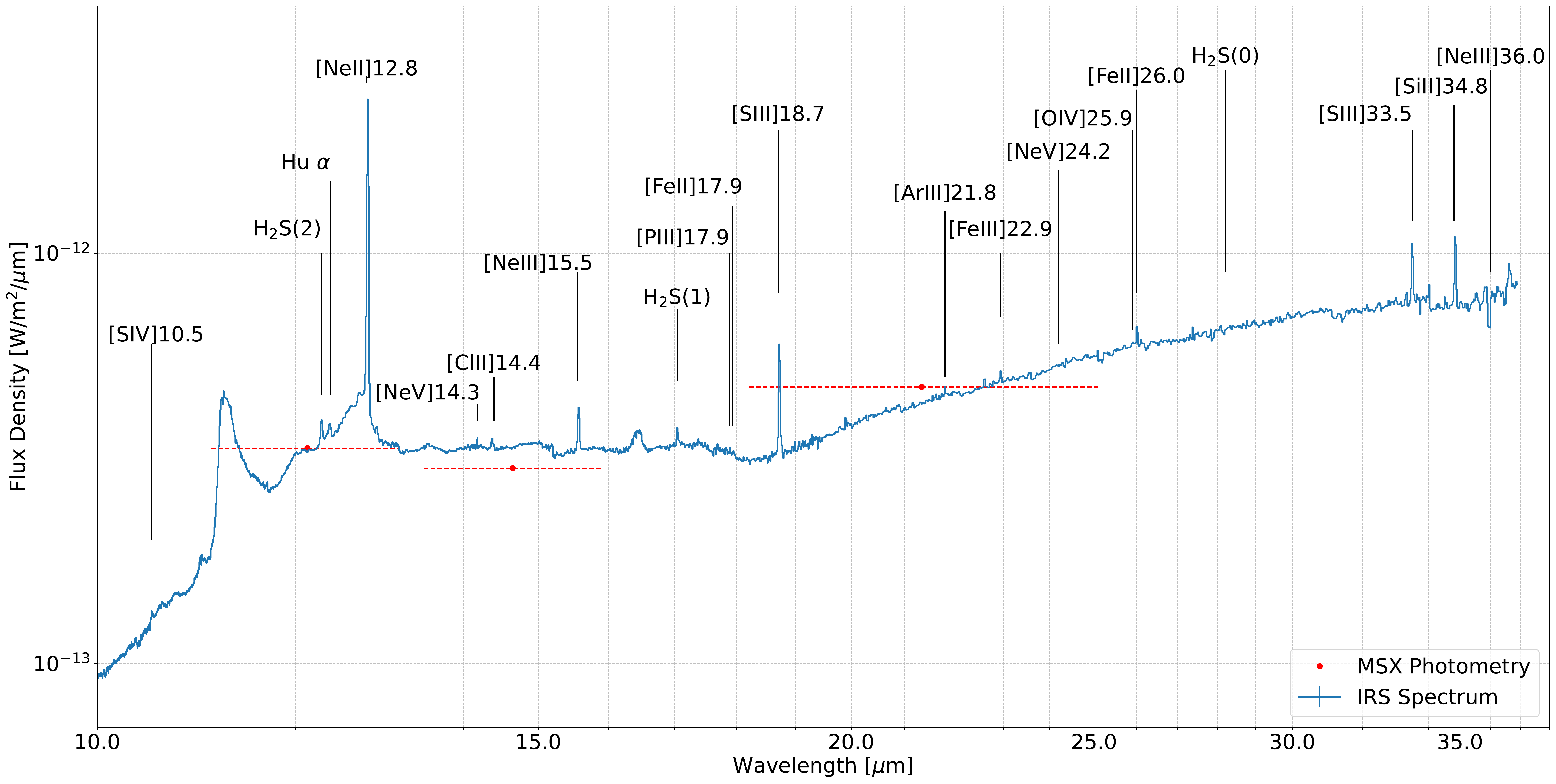}
\caption{\textit{Spitzer}/IRS spectrum (blue) and MSX broadband flux density measurements (red dots) of the nuclear region ($11.1\arcs \times 22.3\arcs$) of NGC\,253. Detected emission lines in the \textit{Spitzer} spectrum are labelled accordingly. MSX filter bandwidths are indicated in dashed lines.}
\label{fig:irsspec}
\end{figure*}

As the spatial extent of the source is of the order of the size of the SH aperture size, the flux density in the SH observations is naturally underestimated due to a loss of flux outside the aperture.
This leads to an offset between the continuum flux density of both, SH and LH modules.
To account for any lost flux density in the smaller SH aperture, we stitched the overlapping continua of both modules.
We fit the spectrum of both modules in the overlapping continuum region ($\sim 18.7 - 19.6\micron$) linearly.
From the fit we get an offset of the two continua of $1.55$, hence, we scale the SH spectrum by a factor of $1.55$.
We assume that the continuum and line emission arise from the same region, which is valid because the MIR continuum emission is mostly from hot dust in $\ion{H}{ii}$ regions.
We used broad band photometric images from the Midcourse Space Experiment (MSX\footnote{available at \url{https://irsa.ipac.caltech.edu/data/MSX/}}) at $\lambda = 12.13\micron, \, 14.65\micron, \, 21.3\micron$ to assert that no wavelength-dependent scaling is necessary.
The more recent WISE observations are saturated and thus cannot be used to verify the scaling.
The continuum of the IRS spectra agrees within $\sim 10\%$ with all three MSX bands, thus no wavelength-dependent scaling is needed.
The resulting spectrum and MSX measurements are shown in Fig. \ref{fig:irsspec}.
Footprints of SH and LH observations can be seen in Fig. \ref{fig:apertures}.

Line fluxes from both modules and other parameters of the observed lines are listed in Table \ref{tab:spitzerIRS}. 
Our values are in good agreement with values from \citet{Bernard-Salas2009} who used the same data set.

\subsubsection{\textit{Herschel}/PACS}\label{sssec:pacs_hom}

To get the integrated line flux from the nuclear source from the PACS line flux maps created in Sect. \ref{sssec:PACS_fit}, we summed up the line flux from the central $3\times 3$ spaxels ($28.2\arcs \times 28.2\arcs$) to limit background emission and noise from the weakly illuminated outer pixels, and to only extract the nuclear source flux.
A comparison of our line fluxes with \citet{PerezBeaupuits2018} shows consistency within the error bars for all lines except $\left[\ion{C}{ii}\right]158\micron$, which is $\sim 20\%$ weaker in our case.
However, when comparing the $\left[\ion{C}{ii}\right]158\micron$ line flux from all $5\times 5$ spaxels, the results are again in good agreement.
The large difference between these two values is due to a significant contribution of $\left[\ion{C}{ii}\right]158\micron$ from the underlying disk.
Line fluxes and errors of the brightest fine-structure lines observed with PACS are listed in Table \ref{tab:spitzerIRS}.

\begin{table*}
\centering
\caption{Line fluxes and uncertainties for the nuclear region, as well as properties of the emission lines observed with \textit{Spitzer}/IRS (Fig. \ref{fig:irsspec}), SOFIA/FIFI-LS (Fig. \ref{fig:fifiFlux}), and \textit{Herschel}/PACS and SPIRE. See Sects. \ref{ssec:FIFILS} and \ref{ssec:ancillaryData} for data reduction, and Sect. \ref{ssec:homogenisation} for extraction of the nuclear emission.}
\begin{tabular}{llrrrr}
\toprule
\toprule
Species & Transition & \multicolumn{1}{c}{Rest Wavelength} & \multicolumn{1}{c}{Flux}  & Critical Density & \multicolumn{1}{c}{Ionisation Energy}\\
	& 			&	\multicolumn{1}{c}{[$\mu$m]} & [$10^{-16}\,\rm{W \, m}^{-2}$] & [$\rm{cm}^{-3}$] & \multicolumn{1}{c}{[eV]} \\
\midrule
\multicolumn{2}{l}{\textit{Spitzer/IRS}} & & & \\
$\left[\ion{S}{iv}\right]$ & $^{2}P_{3/2}$ - $^{2}P_{1/2}$ & 10.511 & $1.29 \pm 0.21$ & $5 \times 10^{4}$ [e] & 34.79 \\
H$_{2}$ & $0-0$ S$(2)$ & 12.279 & $11.6 \pm 1.06$ & $2 \times 10^{5}$ [H$_{2}$] & - \\
Hu $\alpha$ & $n=7-6$ & 12.372 & $6.19 \pm 1.58$ & - & - \\
$\left[\ion{Ne}{ii}\right]$ & $^{2}P_{1/2}$ - $^{2}P_{3/2}$ & 12.814 & $446.40 \pm 44.64$ & $7 \times 10^{5}$ [e] & 21.56 \\
$\left[\ion{Ne}{v}\right]$ & $^{3}P_{2}$ - $^{3}P_{1}$ & 14.322 & $< 0.78$ & $3 \times 10^{4}$ [e] & 97.12 \\
$\left[\ion{Cl}{ii}\right]$ & $^{3}P_{1}$ - $^{3}P_{2}$ & 14.368 & $5.52 \pm 1.06$ & $4 \times 10^{4}$ [e] & 12.97 \\ 
$\left[\ion{Ne}{iii}\right]$ & $^{3}P_{1}$ - $^{3}P_{2}$ & 15.555 & $29.91 \pm 2.33$ & $3 \times 10^{5}$ [e] & 40.96 \\
H$_{2}$ & $0-0$ S$(1)$ & 17.035 & $11.93 \pm 1.19$ & $2 \times 10^{4}$ [H$_{2}$] & - \\
$\left[\ion{P}{iii}\right]$ & $^{2}P_{3/2}$ - $^{2}P_{1/2}$ & 17.885 & $<2.02$ &  & 19.77 \\ 
$\left[\ion{Fe}{ii}\right]$ & $^{7}D_{7/2}$ - $^{7}D_{9/2}$ & 17.936 & $<3.10$ &  & 7.90 \\ 
$\left[\ion{S}{iii}\right]$ & $^{3}P_{2}$ - $^{3}P_{1}$ & 18.713 & $97.81 \pm 3.78$ & $2 \times 10^{4}$ [e] & 23.34 \\
$\left[\ion{Ar}{iii}\right]$ & $^{3}P_{0}$ - $^{3}P_{1}$ & 21.831 & $< 2.00$ & $4 \times 10^{4}$ [e] & 27.63 \\
$\left[\ion{Fe}{iii}\right]$ & $^{5}D_{3}$ - $^{5}D_{4}$ & 22.925 & $10.30 \pm 2.23$ & $10^{5}$ [e] & 16.19 \\
$\left[\ion{Ne}{v}\right]$ & $^{3}P_{1}$ - $^{3}P_{0}$ & 24.318 & $<1.50$ & $3 \times 10^{5}$ [e] & 97.12 \\
$\left[\ion{O}{iv}\right]$ & $^{2}P_{3/2}$ - $^{1}P_{1/2}$ & 25.913 & $7.75 \pm 4.70$ & $1 \times 10^{4}$ [e] & 54.94 \\
$\left[\ion{Fe}{ii}\right]$ & $^{6}D_{7/2}$ - $^{6}D_{9/2}$ & 25.988 & $21.00 \pm 5.15$ & $2 \times 10^{6}$ [H], $10^{4}$ [e] & 7.90 \\
H$_{2}$ & $0-0$ S$(0)$ & 28.221 & $<5.00$ & $700$ [H$_{2}$] & - \\
$\left[\ion{S}{iii}\right]$ & $^{3}P_{1}$ - $^{3}P_{0}$ & 33.480 & $166.00 \pm 19.00$ & $7 \times 10^{3}$ [e] & 23.34 \\
$\left[\ion{Si}{ii}\right]$ & $^{2}P_{3/2}$ - $^{2}P_{1/2}$ & 34.814 & $241.00 \pm 22.40$ & $3 \times 10^{5}$ [H], $10^{3}$ [e] & 8.15 \\
$\left[\ion{Ne}{iii}\right]$ & $^{3}P_{0}$ - $^{3}P_{1}$ & 36.014 & $<20.00$ & $6 \times 10^{4}$ [e] & 40.96 \\
\midrule
\multicolumn{2}{l}{\textit{SOFIA/FIFI-LS}} & & & \\
$\left[\ion{O}{iii}\right]$ & $^{3}P_{2}$ - $^{3}P_{1}$ & 51.815 & $43.34 \pm 4.17$ & $4 \times 10^{3}$ [e] & 35.12 \\
$\left[\ion{O}{iii}\right]$ & $^{3}P_{1}$ - $^{3}P_{0}$ & 88.356 & $87.60 \pm 16.91$ & 510 [e] & 35.12 \\
$\left[\ion{O}{i}\right]$ & $^{3}P_{0}$ - $^{3}P_{1}$ & 145.525 & $43.01 \pm 3.72$ & $10^{5}$ [H] & - \\
$\left[\ion{C}{ii}\right]$ & $^{3}P_{3/2}$ - $^{3}P_{1/2}$ & 157.741 & $452.95 \pm 13.51$ & $3 \times 10^{3}$ [H], 50 [e]  & 11.26 \\
\midrule
\multicolumn{2}{l}{\textit{Herschel/PACS}} & & & \\
$\left[\ion{N}{iii}\right]$ & $^{2}P_{3/2}$ - $^{2}P_{1/2}$ & 57.32 & $57.76 \pm 40.73$ & $3 \times 10^{3}$ [e] & 29.60 \\
$\left[\ion{O}{i}\right]$ & $^{3}P_{1}$ - $^{3}P_{2}$ & 63.185 & $372.50 \pm 57.59$ & $5 \times 10^{5}$ [H] & - \\
$\left[\ion{O}{iii}\right]$ & $^{3}P_{1}$ - $^{3}P_{0}$ & 88.356 & $74.25 \pm 17.19$ & 510 [e] & 35.12 \\
$\left[\ion{N}{ii}\right]$ & $^{3}P_{2}$ - $^{3}P_{1}$ & 121.8 & $94.16 \pm 22.28$ & $300$ [e] & 14.53 \\
$\left[\ion{O}{i}\right]$ & $^{3}P_{0}$ - $^{3}P_{1}$ & 145.525 & $43.06 \pm 7.104$ & $10^{5}$ [H] & - \\
$\left[\ion{C}{ii}\right]$ & $^{3}P_{3/2}$ - $^{3}P_{1/2}$ & 157.741 & $380.30 \pm 44.61$ & $3 \times 10^{3}$ [H], 50 [e]  & 11.26 \\
\multicolumn{2}{l}{\textit{Herschel/SPIRE}} & & & \\
$\left[\ion{N}{ii}\right]$ & $^{3}P_{1}$ - $^{3}P_{0}$ & 205.3 & $20.63 \pm 2.13$ & $45$ [e] & 14.53 \\
$\left[\ion{C}{i}\right]$ & $^{3}P_{2}$ - $^{3}P_{1}$ & 370.415 & $12.25 \pm 1.25$ & $2\times 10^{3}$ [H] & - \\
$\left[\ion{C}{i}\right]$ & $^{3}P_{1}$ - $^{3}P_{0}$ & 609.135 & $4.93 \pm 0.56$ &  $300$ [H] & - \\
\bottomrule
\end{tabular}
\tablefoot{Critical densities from \cite{Malhotra2001}, \cite{Giveon2002}, \cite{Kaufman2006}, \cite{Meijerink2007} and \cite{Cormier2012}, calculated at $T=100$, $300$ and $10000\, \rm{K}$ for collisions with H atoms, H$_{2}$ molecules and electrons, respectively. The critical densities are listed for the main collisional partners. Uncertainties listed are only statistical errors from the fit. In addition, systematic uncertainties are $15\%$ for IRS \citep{Sloan2015} and PACS \citep{PerezBeaupuits2018}, and $20\%$ for FIFI-LS (FIFI-LS team, private communication).  Transitions and wavelengths from NIST \citep{NIST}, ionisation potentials to create the ion from \citet{Draine2011}.}
\label{tab:spitzerIRS}
\end{table*}


\section{Analysis}
We used ratios of our MIR and FIR integrated line fluxes to determine parameters of the ISM.
These fine-structure lines are in most cases not affected by extinction or self-absorption.
Furthermore, when studying line ratios from the same species (e.g. $\left[\ion{O}{iii}\right]52/88\,\micron$) the ratio is independent of the ionic abundance of O$^{2+}$.
This is particularly important, since the elemental abundances are uncertain for NGC\,253.

\subsection{Optical depth}\label{ssec:opticalDepth}
Luminous infrared objects tend to be highly extincted sources, making it sometimes necessary to correct for extinction even at MIR wavelengths, particularly for $A_{V} \geq 10\,\mathrm{mag}$ \citep{Armus2007}.
Hence, we have to correct the MIR line fluxes for extinction or make sure that such correction is not needed for our modelling approach.
The optical depth determinations for the nuclear region in NGC\,253 vary within the literature from $A_{V}=5.5\,\mathrm{mag}$ \citep[][from dust SED fit]{PerezBeaupuits2018}, $A_{V}=14.0\,\mathrm{mag}$ \citep[][from Br$\alpha$/Br$\gamma$]{Rieke1980} up to $A_{V}=19.1\,\mathrm{mag}$ (\citealp{Engelbracht1998}, from $\left[\ion{Fe}{ii}\right]$), depending on the tracers.

To determine the optical depth of the nuclear region of NGC\,253, we used the SED fitting tool MAGPHYS \citep{DaCunha2008}.
Based on a set of photometric observations between the UV and FIR range as input, MAGPHYS fits the SED of a galaxy.
MAGPHYS varies different parameters for the stellar emission (e.g. SFR and initial mass function), and dust attenuation and emission (e.g. optical depth and dust mass) and fits the SED by minimising the $\chi^{2}$ between observations and modelled SED.
It is not possible to fix any of the parameters, all variables are free.
As input data we used photometric observations from the archives of different observatories, covering the spectral range from the UV to the FIR.
We used data from GALEX in the FUV and NUV bands, 2MASS in the $J$, $H$ and $K_{s}$ bands, \textit{Spitzer}/IRAC bands $1-4$ and \textit{Herschel}/PACS bands $70$, $100$, and $160\micron$.
The MSX filters are not available in MAGPHYS and hence are not used to model the SED.
Details about the different photometric bands are reported in Table \ref{tab:magphysPhotBands}.

\begin{table}
    \centering
    \caption{Photometric bands, isophotal wavelengths $\lambda_{\mathrm{Isophot}}$, and flux densities $F$ of the different observatories used for the SED fit with MAGPHYS (Sect. \ref{ssec:opticalDepth}).}
    \begin{tabular}{llrr}
        \toprule
        \toprule
        Observatory & Band & $\lambda_{\mathrm{Isophot}}$ [$\micron$] & $F$ [Jy]\\
        \midrule
        GALEX                  & FUV     & $0.153$                    & $0.03$ \\
                               & NUV     & $0.253$                    & $0.10$ \\
        2MASS                  & $J$     & $1.2$                      & $0.33$ \\
                               & $H$     & $1.65$                     & $0.60$ \\
                               & $K_{s}$ & $2.2$                      & $0.73$ \\
        \textit{Spitzer}/IRAC  & $1$     & $3.6$                      & $0.78$ \\
                               & $2$     & $4.5$                      & $0.87$ \\
                               & $3$     & $5.8$                      & $3.10$ \\
                               & $4$     & $8.0$                      & $8.83$ \\
        \textit{Herschel}/PACS & blue    & $70.0$                     & $779.62$ \\
                               & green   & $100.0$                    & $822.10$ \\
                               & red     & $160.0$                    & $551.37$ \\
        \bottomrule
    \end{tabular}
    \label{tab:magphysPhotBands}
\end{table}

\begin{figure}
    \centering
       \resizebox{\hsize}{!}{\includegraphics{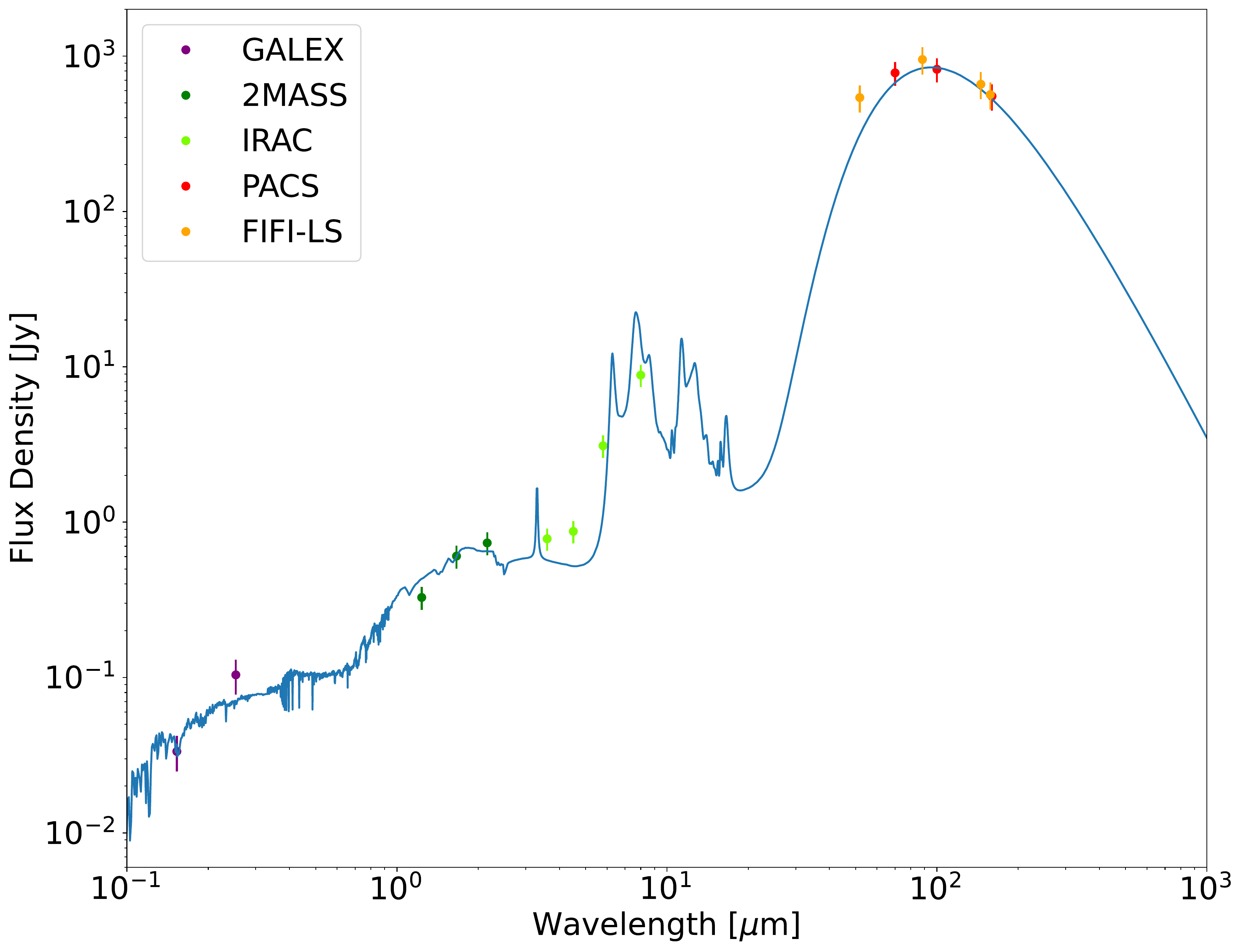}}
    \caption{Broadband measurements from GALEX, 2MASS, \textit{Spitzer}/IRAC, and \textit{Herschel}/PACS observations and the resulting SED fit from MAGPHYS (blue). We also include the continuum from the SOFIA/FIFI-LS observations (Sect. \ref{ssec:FIFILS}) which are in good agreement with the fitted SED.}
    \label{fig:magphys_sedfit}
\end{figure}

We obtained the flux density maps from the respective archives of the different observatories.
To get only the flux density from the nucleus, we convolved the source size of $6.68\arcs$ (see Fig. \ref{fig:LH_spatialprofile}) with the resolution of the different observatories at the given wavelength and extracted the flux density from an aperture of the resulting diameter.
We assume that the nuclear emission is dominant in this region and that emission from other structures like the bar are negligible.
The obtained measurements are listed in Table \ref{tab:magphysPhotBands} and shown in Fig. \ref{fig:magphys_sedfit}.
Using these flux densities as input, MAGPHYS determines the best-fitting SED and returns the fitted parameters.
From the fit, we obtained a total infrared luminosity $L_{\mathrm{TIR}} = 1.37\times 10^{10}\,\mathrm{L}_{\odot}$.
We also used the prescription given in \citet{Galametz2013} (see their Eq. 7 and Table 3) to calculate $L_{\mathrm{TIR}}$ from FIR data only. 
From this prescription - using the \textit{Herschel}/PACS photometric data only - we obtain $L_{\mathrm{TIR}} = (9.2 \pm 0.4) \times 10^{9} \,L_{\odot}$, which is in good agreement with the result from MAGPHYS.
MAGPHYS also returns the optical depth, which is $A_{V} = 4.35\,\mathrm{mag}$.
This again is in good agreement with a recent study \citep{PerezBeaupuits2018} who obtained $A_{v} = 5.5 \pm 2.5\,\mathrm{mag}$ from the $100\micron$ dust continuum emission.
The fitted SED is shown in Fig. \ref{fig:magphys_sedfit}.

Using the models from \citet{Weingartner2001} with $R_V=3.1$ and the optical depth we obtained ($A_{V}=4.35\,\mathrm{mag}$) we calculated optical depths at other wavelengths.
We yield an optical depth at $\lambda=20\micron$ of $A_{20\micron} = 0.13\,\mathrm{mag}$, which translates into an extinction correction for the line fluxes of $\sim 12\%$.
The wavelengths of the emission lines used for the line flux ratios are close, thus, the extinction affects the line ratios even less.
In case of the $(\left[\ion{Ne}{ii}\right]13\micron + \left[\ion{Ne}{iii}\right]16\micron)/$Hu $\alpha$, the extinction correction is $3.5\%$.
We find that the extinction correction has little effect for the set of lines used in our study.

\subsection{Ionised gas density}\label{ssec:cloudyDensity}
Densities of the nuclear region of NGC\,253 have been reported in the literature to range from $\sim 4.3 \times 10^{2} \, \rm{cm}^{-3}$ (\citealp{Carral1994}, from $\left[\ion{O}{iii} \right]$) up to $5\times 10^{3} \, \rm{cm}^{-3}$ (\citealp{Engelbracht1998}, from $\left[\ion{Fe}{ii} \right]$) for the ionised gas.
The results depend on the tracers used which could, in principle, apply to different regions within the observing beam or could be associated with different regions due to poorer spatial resolution.

\begin{figure*}
    \centering
    \includegraphics[width=17cm]{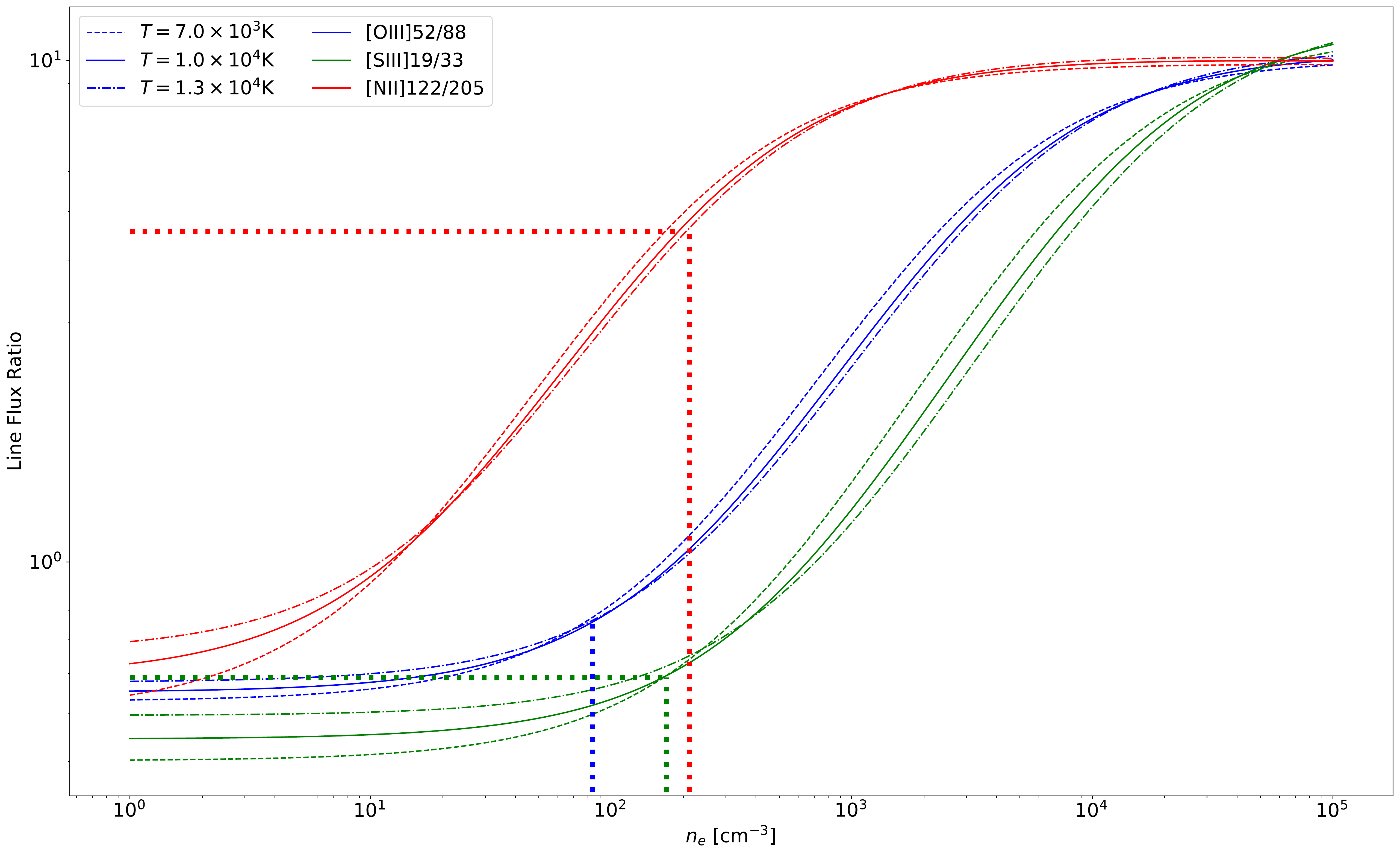}
    \caption{Density dependence (logscale) of the $\left[\ion{O}{iii}\right]52\micron$/$\left[\ion{O}{iii}\right]88\micron$, $\left[\ion{N}{ii}\right]122\micron$/$\left[\ion{N}{ii}\right]205\micron$, and $\left[\ion{S}{iii}\right]19\micron$/$\left[\ion{S}{iii}\right]33\micron$ line flux ratios for different temperatures. The various line flux ratios are sensitive to different density regimes. The results from $\left[\ion{N}{ii}\right]122\micron$/$\left[\ion{N}{ii}\right]205\micron$ and $\left[\ion{S}{iii}\right]19\micron$/$\left[\ion{S}{iii}\right]33\micron$ are indicated in red and green dots respectively, for $\left[\ion{O}{iii}\right]52\micron$/$\left[\ion{O}{iii}\right]88\micron$ we show the upper density limit (blue dots).}
    \label{fig:densityRanges}
\end{figure*}

Several observed line flux ratios are sensitive to the density of the \ion{H}{ii} region.
Well known examples are the $\left[\ion{O}{iii}\right]52\micron$/$\left[\ion{O}{iii}\right]88\micron$ and $\left[\ion{N}{ii}\right]122\micron$/$\left[\ion{N}{ii}\right]205\micron$ ratios \citep[e.g. ][]{Carral1994,Oberst2011,HerreraCamus2016}.
Additionally, the MIR line flux ratios $\left[\ion{Ne}{iii}\right]16\micron$/$\left[\ion{Ne}{iii}\right]36\micron$, $\left[\ion{Ne}{v}\right]14\micron$/$\left[\ion{Ne}{v}\right]24\micron$, and $\left[\ion{S}{iii}\right]19\micron$/$\left[\ion{S}{iii}\right]33\micron$ are excellent tracers for the density of the ionised gas, since they are independent of the ionic abundance and weakly dependent on the electron temperature.
Each of these line flux ratios are not only ideal tracers of the density, due to their different critical densities (see Table \ref{tab:spitzerIRS}), they potentially also trace different regions of the ionised ISM, as shown in Fig. \ref{fig:densityRanges}.
Owing to the critical densities, the line flux ratio $\left[\ion{N}{ii}\right]122\micron$/$\left[\ion{N}{ii}\right]205\micron$ is a good tracer for the diffuse outskirts of $\ion{H}{ii}$ regions \citep{Oberst2011}, the ratios $\left[\ion{S}{iii}\right]19\micron$/$\left[\ion{S}{iii}\right]33\micron$ and $\left[\ion{Ne}{iii}\right]16\micron$/$\left[\ion{Ne}{iii}\right]36\micron$ trace a higher density phase.
Furthermore, the $\left[\ion{Ne}{v}\right]14\micron$/$\left[\ion{Ne}{v}\right]24\micron$ traces the gas phase whose conditions are dominated by the AGN, since Ne$^{4+}$ is not produced by stars but by the accretion disk from an AGN or fast radiative shocks \citep{Izotov2012}.

We used a Monte-Carlo method to obtain a density from these line flux ratios and also to propagate uncertainties.
With the Monte-Carlo method we were also able to investigate if the influence of the electron temperature is indeed negligible. 

To determine the hydrogen density, we used the Python package \textit{PyNeb} \citep{Luridiana2013}.
\textit{PyNeb} calculates the density of the ionised gas, if two density dependent emission lines and the temperature of the gas are provided, by solving the equilibrium equations of the $n$-level atom and determining the level populations.
We assumed a normal distribution for the line fluxes, where the mean of the distribution is given by the line flux and the standard deviation corresponds to the statistical uncertainties given in Table \ref{tab:spitzerIRS}.
For the calculations, the temperature is uniformly distributed between $7000\,\mathrm{K}$ and $13000\,\mathrm{K}$, which are typical values for the gas in $\ion{H}{ii}$ regions \citep{Osterbrock2006}.
In each of the $10^{5}$ steps, a temperature and line flux for each of the ionised species of a pair of lines were randomly picked and a density was calculated.
The results of these calculations are shown in the left panel of Fig. \ref{fig:MC_NeonAbundance_Sulfur} for the $\left[\ion{S}{iii}\right]19\micron$/$\left[\ion{S}{iii}\right]33\micron$ line flux ratio. 

We applied this Monte-Carlo method to all line flux ratios listed above.
However, from the $\left[\ion{Ne}{iii}\right]16\micron$/$\left[\ion{Ne}{iii}\right]36\micron$ and $\left[\ion{Ne}{v}\right]14\micron$/$\left[\ion{Ne}{v}\right]24\micron$ we did not obtain any results as these line flux ratios did not allow \textit{PyNeb} to solve the level populations.
We derived the densities listed in Table \ref{tab:results_densities} from the line flux ratios.

\begin{table}[]
    \centering
    \caption{Densities derived from the line flux ratios as discussed in Sect. \ref{ssec:cloudyDensity}}
    \begin{tabular}{lll}
        \toprule
        \toprule
        Lines & Line Flux Ratio  & Density [cm$^{-3}$] \\
        \midrule
        $\left[\ion{S}{iii}\right]19/33\micron$ & $0.59$ & $170.0 \pm 83.6$ \\
        $\left[\ion{O}{iii}\right]52/88\micron$ & $0.49$ & $<83.6$ \\
        $\left[\ion{N}{ii}\right]122/205\micron$ & $4.56$ & $211.7 \pm 105.6$ \\
        \bottomrule
    \end{tabular}
    \tablefoot{The $\left[\ion{O}{iii}\right]$ yields densities in the low density regime, where the line ratio is not sensitive to the density, hence we give only an upper limit. The $\left[\ion{Ne}{v}\right]14\micron$, $\left[\ion{Ne}{v}\right]24\micron$, and $\left[\ion{Ne}{iii}\right]36\micron$ are upper limits and hence not suitable to determine a density.}
    \label{tab:results_densities}
\end{table}

The obtained densities from $\left[\ion{S}{iii}\right]19\micron$/$\left[\ion{S}{iii}\right]33\micron$ and $\left[\ion{N}{ii}\right]122\micron$/$\left[\ion{N}{ii}\right]205\micron$ are in good agreement with each other despite the different critical densities.
In contrast, the $\left[\ion{O}{iii}\right]52\micron$/$\left[\ion{O}{iii}\right]88\micron$ is lower by a factor of $2$.
The density obtained from the ratio $\left[\ion{O}{iii}\right]52\micron$/$\left[\ion{O}{iii}\right]88\micron$ lies within a range where the line flux ratio is not sensitive to the density, hence, we only give an upper limit for the density.
For the latter line flux ratio, we used both lines from the FIFI-LS observations to reduce systematic effects.

The densities derived from all line flux ratios are significantly lower than results from other studies \citep[e.g. ][]{Engelbracht1998}.
For the $\left[\ion{N}{ii}\right]$ line flux ratio, this is most likely due to the low critical densities of both transitions.
Both, the $\left[\ion{N}{ii}\right]122\micron$ and $\left[\ion{N}{ii}\right]205\micron$ lines are hence associated with diffuse, low density regions and do not trace high density regimes.
The $\left[\ion{S}{iii}\right]$ line flux ratio traces higher density regions (see Fig. \ref{fig:densityRanges}), but the S$^{+}$ ion has a higher ionisation potential (see Table \ref{tab:spitzerIRS}) compared to, for example, Fe$^{+}$.
Fe$^{+}$, however was used in \citet{Engelbracht1998}. 
The different excitation potential and critical density of the species could explain the discrepancy between their and our study. 
The ratio $\left[\ion{O}{iii}\right]52\micron$/$\left[\ion{O}{iii}\right]88\micron$ was also used by \citet{Carral1994}.
The different results from \citet{Carral1994} and this study is most likely due to the different sizes of observing beams, leading to a higher $\left[\ion{O}{iii}\right]52\micron$ line flux and, hence, to a higher density, as discussed in Sect. \ref{sssec:fifi_hom}.
We conclude that, although all densities are in moderate agreement with each other, to appropriately model the ionised gas in the nuclear region at least a two phase model would be needed, due to the different critical densities and ionisation potentials of the used species.

\subsection{Chemical composition}\label{ssec:chemicalComp}
The determinations of metallicity in the nuclear region of NGC\,253 vary widely within literature.
The metallicity ranges from $\sim 0.5\,Z_{\odot}$ \citep{Carral1994,Ptak1997}, $\lesssim 1\,Z_{\odot}$ \citep{Engelbracht1998}, up to $\lesssim 1.5\,Z_{\odot}$ \citep{Webster1983}.

We calculated the metallicity with the available emission lines, self-consistently.
The $\left[\ion{Ne}{ii}\right]13\micron$, $\left[\ion{Ne}{iii}\right]16\micron$, and Hu $\alpha$ together provide excellent constraints for this purpose, by using the $(\left[\ion{Ne}{ii}\right]13\micron + \left[\ion{Ne}{iii}\right]16\micron)/$Hu $\alpha$ line flux ratio, which is sensitive to the Ne/H abundance in the gas \citep{Bernard-Salas2001,Lebouteiller2008}.
The lines arise from the ionised gas and Ne is not depleted on dust grains, hence, the Ne/H is an ideal tracer for the metallicity of the gas.

To calculate the Ne/H abundance ratio, we again used the Python tool \textit{PyNeb}.
For a given temperature $T$, density $n$, and line flux ratio $\left[\ion{Ne}{ii}\right]13\micron$/Hu $\alpha$, \textit{PyNeb} predicts the Ne$^{+}$/H abundance ratio by solving the equilibrium equations and level populations for the Ne and H atom.
The procedure is the same for the line flux ratio $\left[\ion{Ne}{iii}\right]16\micron$/Hu $\alpha$, yielding the abundance ratio Ne$^{2+}$/H.
We assumed that Ne$^{+}$ and Ne$^{2+}$ are the dominant ionisation stages, that is $\mathrm{Ne/H} = \mathrm{Ne}^{+}/\mathrm{H} + \mathrm{Ne}^{2+}/\mathrm{H}$.
Under this assumption we can use the densities derived in Sect. \ref{ssec:cloudyDensity}.

Again, we used the Monte-Carlo method to propagate uncertainties in the derived Ne/H abundance ratio.
We calculated the Ne/H abundance ratio assuming the different densities we obtained from the $\left[\ion{O}{iii}\right]$, $\left[\ion{N}{ii}\right]$, and $\left[\ion{S}{iii}\right]$ line flux ratios in Sect. \ref{ssec:cloudyDensity}.
To do so we extended our Monte-Carlo approach from Sect. \ref{ssec:cloudyDensity}.
After calculating the density from the randomly drawn line fluxes (e.g. for $\left[\ion{S}{iii}\right]19\micron$ and $\left[\ion{S}{iii}\right]33\micron$), line fluxes for $\left[\ion{Ne}{ii}\right]13\micron$, $\left[\ion{Ne}{iii}\right]16\micron$, and Hu $\alpha$ were drawn $-$ the distributions for these lines are again normal distributions.
With these three emission lines, as well as the temperature and density from the first step (Sect. \ref{ssec:cloudyDensity}), we calculated the Ne/H abundance ratio for $10^{5}$ steps.
From this sample we calculated the mean and standard deviation.

\begin{figure*}
    \centering
    \includegraphics[width=17cm]{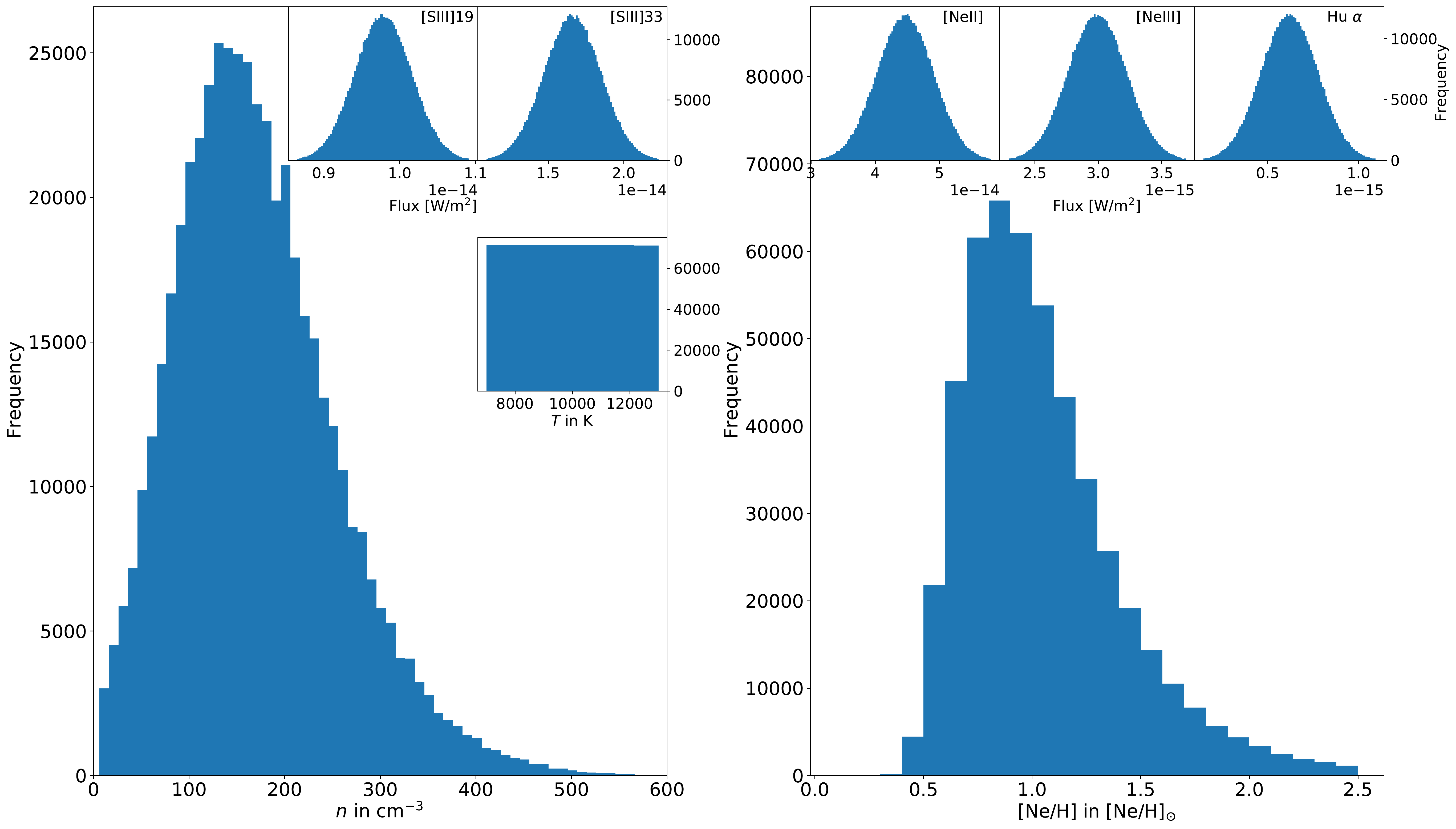}
    \caption{Results from the Monte Carlo simulation to derive the Hydrogen density (Sect. \ref{ssec:cloudyDensity}) and the Ne/H abundance ratio (Sect. \ref{ssec:chemicalComp}). \textit{Left:} The main plot shows the derived density in $10\,\mathrm{cm}^{-3}$ bins. Inset plots show histograms of the randomly picked temperature (uniform distribution, $1000\,\mathrm{K}$ bins), and integrated line fluxes for $\left[\ion{S}{iii}\right]19\micron$ and $\left[\ion{S}{iii}\right]33\micron$.  \textit{Right:} Inset plots again show histograms of the randomly picked integrated line fluxes of $\left[\ion{Ne}{ii}\right]13\micron$, $\left[\ion{Ne}{iii}\right]16\micron$, and Hu $\alpha$. From the line fluxes, temperature and density we obtained the Ne/H abundance ratio in solar units, which is shown in the main plot ($0.1$ bins).}
    \label{fig:MC_NeonAbundance_Sulfur}
\end{figure*}

The results are shown in the right panels of Fig. \ref{fig:MC_NeonAbundance_Sulfur}.
We obtain a Ne/H abundance ratio of nearly solar value, $\left[\mathrm{Ne/H}\right]_{\left[\ion{O}{iii}\right]} = 1.08 \pm 0.41$, $\left[\mathrm{Ne/H}\right]_{\left[\ion{N}{ii}\right]} = 1.00 \pm 0.24$, and $\left[\mathrm{Ne/H}\right]_{\left[\ion{S}{iii}\right]} = 1.00 \pm 0.24$.
The similar results suggest that the density does not seem to play a major role in the metallicity determination in the examined density regimes.
From the Ne/H abundance, we use equations given by \citet{Izotov1999,Izotov2006,Nicholls2017}, who studied elemental abundances in galaxies, to deduce abundances for C, N, O, Si, S, and Fe.
The abundances of these elements are given in Table \ref{tab:abundances}.

\begin{table}
\centering
\caption{Gas-phase elemental abundances relative to the hydrogen abundance for most of the species that were observed.}
\begin{tabular}{lrr}
\toprule
\toprule
Species & Abundance (log) & Ref \\
\midrule
Ne/H & $-4.000$ & \\
\midrule
C/H & $-3.696$ & (3) \\
N/H & $-4.341$ & (3) \\
O/H & $-3.313$ & (2) \\
Si/H & $-4.763$ & (1) \\
S/H & $-5.053$ & (2) \\
Fe/H & $-5.583$ & (2) \\
\bottomrule
\end{tabular}
\tablefoot{Abundances were calculated from the Ne/H abundance (see Sect. \ref{ssec:chemicalComp}) and by the use of equations from \citet[][1]{Izotov1999}, \citet[][2]{Izotov2006}, and \citet[][3]{Nicholls2017}.}
\label{tab:abundances}
\end{table} 

As discussed in, for instance, \citet{Madden2020}, the metallicity is of great importance for determining the CO-to-H$_{2}$ conversion factor $\alpha_{\mathrm{CO}}$.
Using the results from previous studies, $\alpha_{\mathrm{CO}}$ (in $\mathrm{M}_{\odot}\,\mathrm{pc}^{-2}\,\left(\mathrm{K\,km\,s^{-1}}\right)^{-1}$) ranges from $\sim 1$ up to $\sim 40$, covering more than $1.5$ orders of magnitude.
From our obtained metallicity and Eq. (6) in \citet{Madden2020}, we yield a more accurate $\alpha_{\mathrm{CO}}$ of $3.8_{-2.0}^{+5.8}$. 


\section{Summary and outlook}
In this first of a series of papers we presented a wealth of MIR and FIR data of the nuclear region of the starburst galaxy NGC\,253 from the SOFIA, \textit{Herschel}, and \textit{Spitzer} telescopes.
We described the data reduction steps for each observatory and noted caveats that appeared during these processes.
The full list of observed lines, the line fluxes and errors from the nuclear region are listed in Table \ref{tab:spitzerIRS}, the extended line flux maps we obtained from SOFIA/FIFI-LS observations are shown in Fig. \ref{fig:fifiFlux}.

Furthermore, we used the obtained line fluxes to derive fundamental parameters of the local ISM, such as density $n$, metallicity $Z$, and optical depth $A_{V}$.
We showed that for our observed MIR emission lines no extinction correction is needed.
We used the line flux ratios of $\left[\ion{S}{iii}\right]$, $\left[\ion{O}{iii}\right]$, and $\left[\ion{N}{ii}\right]$, which are almost independent of the local temperature and mostly depend on the gas density or, more precisely, on the local density of electrons, which are the main collisional partners to excite these states.
We conclude that there are at least two ionised gas phases in the central region of this galaxy.
From the calculated density and assumed temperature, we used the line flux ratio of $(\left[\ion{Ne}{ii}\right]13\micron + \left[\ion{Ne}{iii}\right]16\micron) / \mathrm{Hu}\, \alpha$ which is an ideal tracer for the Ne/H abundance, to calculate the local metallicity, for which we obtain solar value.

In a forthcoming paper we will use all observed lines to create a more sophisticated multi-phase model using \textit{MULTIGRIS}, a Bayesian approach to determine the probability density distribution within a set of \verb|Cloudy| models \citep{Lebouteiller2022,Ramambason2022} of ISM parameters such as density, intensity of radiation field and age of the starburst.
\begin{acknowledgements}
      The authors thank the anonymous referee for useful comments that improved the quality and clarity of the paper.
      SOFIA, the Stratospheric Observatory For Infrared Astronomy, is a joint project of the Deutsche Raumfahrtagentur im Deutschen Zentrum für Luft- und Raumfahrt e.V. (German Space Agency at DLR, grant: FKZ 50OK2002) and the National Aeronautics and Space Administration (NASA). 
      It is funded on behalf of the German Space Agency at DLR by the Federal Ministry for Economic Affairs and Climate Action based on legislation by the German Parliament, the state of Baden-Württemberg and the University of Stuttgart.
      Scientific operation for Germany is coordinated by the Deutsches SOFIA Institut (DSI) at the University of Stuttgart.
      \\
      \textit{Software:} Astropy \citep{astropy:2013, astropy:2018}, Photutils \citep{photutils2020}, Matplotlib \citep{Matplotlib2007}, NumPy \citep{numpy2020}, SciPy \citep{SciPy2020}
\end{acknowledgements}

\bibliography{Literature}

%
   \bibliographystyle{aa} 
%

\end{document}